# Classification of non conformally flat cylindrically symmetric non static space-times according to their proper conformal motions

Ghulam Shabbir, M. Ramzan and Suhail Khan

Faculty of Engineering Sciences,

GIK Institute of Engineering Sciences and Technology,

Topi, Swabi, NWFP, Pakistan.

Email: shabbir@giki.edu.pk

**Abstract**

In this paper we considered the most general form of non conformally flat cylindrically symmetric non-static space-times to study proper conformal motions using direct integration technique. We have shown that very special classes for cylindrically symmetric space-times admit proper conformal motions. This classification also covers non-static and static plane symmetric space-times. In [9] it was claimed that non conformally flat plane symmetric static space-times do not admit proper conformal motion. Here it is also shown that static and non static plane symmetric space-times admit proper conformal motions. We also discuss the Lie algebra in each case.

## 1. INTRODUCTION

In general relativity theory some exact solutions of Einstein's field equations have been found to understand the physical situation, such as Schwarzchild solution for black hole and Friedmann solutions for cosmology, but due to the non linearity of the field equations it is not always possible to find exact solutions which clearly describe physical situation. Symmetry restrictions are imposed in general relativity to understand the physical and geometrical properties of the space-times and to obtain physically significant solutions of the Einstein's field equations. Therefore much interest has been taken to study symmetries like Killing, homothetic and conformal motions by many authors [1-7]. They adopted different approaches to classify some well known space-times according to their Killing, homothetic and conformal motions. A general discussion of conformal vector fields in space-times is given in [6]. They argued that the maximum dimension of the conformal algebra for space-times that are not conformally flat is seven and for conformally flat space-times the dimension of the conformal algebra is fifteen. D. Kramer et al [5] considered the three dimensional Lie algebra of the rigid rotating, stationary axially symmetric space-times and found that these space-times do not admit proper conformal vector fields which satisfies the Lie algebra structure. P. S. Apostolopoulous et al [7] developed a method for the calculation of the conformal vector field of a general $1+(n-1)$ decomposable metrics. Using their method they confirm that the dimension of the conformal algebra for conformally flat space-times is fifteen and that the Gödel metric does not admit conformal vector fields.

To understand and explore physically significant metrics there was a need to classify other space-times with respect to their conformal vector fields. For this purpose we have discussed previously proper conformal vector fields in different static and non-static space-times [10,11]. In this paper a special emphasis has been given to the non-static cylindrically symmetric space-times for finding conformal motions or conformal vector fields. This classification will also cover static and non static plane symmetric space-times. Throughout in this paper, $M$ represents a four dimensional, connected, Hausdorff space-time manifold with Lorentz metric $g_{ab}$ of signature (-, +, +, +). The Weyl tensor components are denoted in component form $C^a{}_{bcd}$. The usual covariant, partial and Lie derivatives are denoted by a semicolon, a comma and the symbol $L$, respectively. Round and square brackets denote the usual symmetrization and skew-symmetrization, respectively. The space-time $M$ will be assumed non conformally flat in the sense that the Weyl tensor does not vanish over any non empty open subset of $M$.

Any vector field $X$ on $M$ can be decomposed as

$$X_{a;b} = \frac{1}{2}h_{ab} + F_{ab} \tag{1}$$

where $h_{ab}(=h_{ba}) = L_X g_{ab}$ and $F_{ab}(=-F_{ba})$ are symmetric and skew symmetric tensors on $M$, respectively. Such a vector field $X$ is called conformal motion or conformal vector field if the local diffeomorphisms $\phi_t$ (for appropriate $t$) associated with $X$ preserve the metric structure up to a conformal factor i.e. $\phi_t^* g = \psi g$, where $\psi$ is a nowhere zero positive function on some open subset of $M$ and $\phi_t^*$ is a pullback map on some open subset of $M$. This is equivalent to the condition that

$$h_{ab} = \psi g_{ab},$$

equivalently,

$$g_{ab,c} X^c + g_{cb} X^c_{,a} + g_{ac} X^c_{,b} = \psi g_{ab}, \qquad (2)$$

where $\psi : U \to R$ is the smooth conformal function on some subset of $M$, then $X$ is called conformal motion or conformal vector field. If $\psi$ is constant on $M$ then $X$ is homothetic (proper homothetic if $\psi \neq 0$) while if $\psi = 0$ it is Killing [3]. If the vector field $X$ is not homothetic then it is called proper conformal motion.

Further, the conformal vector field $X$ on $M$ also satisfies [3]

$$L_X C^a{}_{bcd} = 0. \qquad (3)$$

The vector field $X$ satisfying the above equation (3) is called a Weyl collineation (WC). The vector field $X$ is called proper WC if it is not conformal [3]. It also follows from [3] that conformal vector field is Weyl collineation but converse does not hold. It follows from [8] that equation (3) is known as the integrability conditions of conformal vector fields. The Lie algebra of a set of vector fields on a manifold is completely characterized by the structure constants $N^a_{bc}$ given in term of the Lie brackets by

$$[X_b, X_c] = N^a_{bc} X_a, \qquad N^a_{bc} = -N^a_{cb},$$

where $X_a$ are the generators and $a,b,c = 0,1,....,n$. The Lie algebra for the conformal vector fields is also discussed in each case.

## 2. Main Results

The most general form of the metric for non static cylindrically symmetric space-time in usual coordinate system $(t,r,\theta,z)$ (labeled by $(x^0, x^1, x^2, x^3)$, respectively) is given by [8]

$$ds^2 = -e^{A(t,r)} dt^2 + dr^2 + e^{B(t,r)} d\theta^2 + e^{C(t,r)} dz^2, \qquad (4)$$

where $A$, $B$ and $C$ are functions of $t$ and $r$ only. The above space-time (4) admits minimum two independent Killing vector fields $\frac{\partial}{\partial \theta}$ and $\frac{\partial}{\partial z}$. A vector field $X$ is said to be conformal if it satisfies equation (2). One can write (2) explicitly by using (4) as

$$\dot{A} X^0 + A' X^1 + 2 X^0{}_{,0} = \psi(t,r,\theta,z) \tag{5}$$

$$X^1{}_{,0} - e^{A(t,r)} X^0{}_{,1} = 0 \tag{6}$$

$$e^{B(t,r)} X^2{}_{,0} - e^{A(t,r)} X^0{}_{,2} = 0 \tag{7}$$

$$e^{C(t,r)} X^3{}_{,0} - e^{A(t,r)} X^0{}_{,3} = 0 \tag{8}$$

$$2 X^1{}_{,1} = \psi(t,r,\theta,z) \tag{9}$$

$$e^{B(t,r)} X^2{}_{,1} + X^1{}_{,2} = 0 \tag{10}$$

$$e^{C(t,r)} X^3{}_{,1} + X^1{}_{,3} = 0 \tag{11}$$

$$\dot{B} X^0 + B' X^1 + 2 X^2{}_{,2} = \psi(t,r,\theta,z) \tag{12}$$

$$e^{C(t,r)} X^3{}_{,2} + e^{B(t,r)} X^2{}_{,3} = 0 \tag{13}$$

$$\dot{C} X^0 + C' X^1 + 2 X^3{}_{,3} = \psi(t,r,\theta,z), \tag{14}$$

where 'dot' denotes the partial derivative with respect to $t$ and 'dash' denote the partial derivative with respect to $r$. Equations (6), (9), (10) and (11) give

$$\begin{aligned} X^0 &= \frac{1}{2}\left(\int e^{-A(t,r)} \int \psi_t(t,r,\theta,z) dr\right) dr + K^1_t(t,\theta,z) \int e^{-A(t,r)} dr + K^2(t,\theta,z), \\ X^1 &= \frac{1}{2} \int \psi(t,r,\theta,z) dr + K^1(t,\theta,z), \\ X^2 &= -\frac{1}{2}\left(\int e^{-B(t,r)} \int \psi_\theta(t,r,\theta,z) dr\right) dr - K^1_\theta(t,\theta,z) \int e^{-B(t,r)} dr + K^3(t,\theta,z), \\ X^3 &= -\frac{1}{2}\left(\int e^{-B(t,r)} \int \psi_z(t,r,\theta,z) dr\right) dr - K^1_z(t,\theta,z) \int e^{-C(t,r)} dr + K^4(t,\theta,z), \end{aligned} \tag{15}$$

where $K^1(t,\theta,z)$, $K^2(t,\theta,z)$, $K^3(t,\theta,z)$ and $K^4(t,\theta,z)$ are functions of integration. In order to find conformal vector fields we need to determine the functions $K^1(t,\theta,z)$, $K^2(t,\theta,z)$, $K^3(t,\theta,z)$ and $K^4(t,\theta,z)$ by integrating the remaining six equations. To avoid lengthy details, here we will present only results, when the above space-times (4) admit conformal vector fields. The following possibilities exist when the above space-times (4) admit conformal vector fields which are:

**Case (1)**

In this case the metric functions for the space-time (4) becomes

$$A(t,r) = (c_8 t + c_9) e^{-c_6 \int V(r) dr} + \ln U^2(r) - 2c_6 \int V(r) dr - c_{10} \int V(r) e^{-c_6 \int V(r) dr} dr,$$

$$B(t,r) = (c_{11} t + c_{12}) e^{-c_6 \int V(r) dr} + \ln U^2(r) - 2c_3 \int V(r) dr - c_{13} \int V(r) e^{-c_6 \int V(r) dr} dr,$$

$$C(t,r) = (c_{11} t + c_{12}) e^{-c_6 \int V(r) dr} + \ln U^2(r) - 2c_1 \int V(r) dr - c_{13} \int V(r) e^{-c_6 \int V(r) dr} dr,$$

where $V(r) = \dfrac{1}{U(r)}$, $U(r) = \dfrac{1}{2} \int \psi(r) dr + c_5$ and $c_1, c_3, c_5, c_6, c_7, c_8, c_9, c_{10}, c_{11},$

$c_{12}, c_{13}, c_{14} \in R (c_6 \neq 0, c_{11} \neq 0, c_8 \neq 0, c_1 \neq c_3, c_{10} = c_7 c_8 - c_6 c_9, c_{13} = c_7 c_{11} - c_6 c_{12})$. The sub case when $c_1 = c_3$ will be consider later. The conformal vector fields in this case are

$$X^0 = c_6 t + c_7, \quad X^1 = U(r) c_{14}, \quad X^2 = \theta c_3 + c_4, \quad X^3 = z c_1 + c_2 \tag{16}$$

and the conformal factor is $\psi(r) = 2 \dfrac{d}{dr} U(r)$. In this case the space-time (4) admits three independent conformal vector fields in which one is proper conformal and two are independent Killing vector fields. It is important to tell the readers that $c_1, c_3, c_6, c_7$ are the part of the space-times hence they can not be the generators. Generators in this case are

$$X_1 = U(r) \frac{\partial}{\partial r}, \quad X_2 = \frac{\partial}{\partial \theta}, \quad X_3 = \frac{\partial}{\partial z}.$$

Here all the structure constants are zero that is $[X_b, X_c] = 0$, for all $b, c = 1, 2, 3$. In this case Lie algebra is closed. Proper conformal vector field after subtracting Killing vector fields is

$$X = (c_6 t + c_7, U(r) c_{14}, \theta c_3, z c_1). \tag{17}$$

Now consider the sub case when $c_1 = c_3$, the above space-time (4) takes the form

$$ds^2 = -e^{A(t,r)} dt^2 + dr^2 + e^{B(t,r)} (d\theta^2 + dz^2), \tag{18}$$

where $A(t,r) = (c_8 t + c_9) e^{-c_6 \int V(r) dr} + \ln U^2(r) - 2c_6 \int V(r) dr - c_{10} \int V(r) e^{-c_6 \int V(r) dr} dr,$

$B(t,r) = (c_{11} t + c_{12}) e^{-c_6 \int V(r) dr} + \ln U^2(r) - 2c_3 \int V(r) dr - c_{13} \int V(r) e^{-c_6 \int V(r) dr} dr$ and

$U(r) = \frac{1}{2}\int \psi(r)dr + c_5$. The above space-time (18) becomes the most general form of the non static plane symmetric space-times and it admits three independent Killing vector fields which are $\frac{\partial}{\partial \theta}, \frac{\partial}{\partial z}, \theta \frac{\partial}{\partial z} - z \frac{\partial}{\partial \theta}$. It also follows from the above calculation that the above space-time (18) admits four conformal vector fields which are

$$X = (c_6 t + c_7, U(r)c_{14}, \theta c_3 - z c_2 + c_4, z c_3 + \theta c_2 + c_8). \qquad (19)$$

Conformal factor in this case is $\psi(r) = 2\frac{d}{dr}U(r)$. It is important to mention here that $c_3, c_6, c_7$ are the part of the space-times hence they can not be the generators. Generators in this case are

$$X_1 = U(r)\frac{\partial}{\partial r}, \quad X_2 = \frac{\partial}{\partial \theta}, \quad X_3 = \frac{\partial}{\partial z}, \quad X_4 = \theta \frac{\partial}{\partial z} - z \frac{\partial}{\partial \theta}.$$

Here the non zero components of the Lie brackets are $[X_2, X_4] = X_3$ and $[X_3, X_4] = -X_2$. In this case Lie algebra is closed. Proper conformal vector field after subtracting Killing vector fields from equation (19) is

$$X = (c_6 t + c_7, U(r)c_{14}, \theta c_3, z c_3). \qquad (20)$$

**Case (2)**

In this case the space-time (4) becomes

$$ds^2 = -e^{A(r)} dt^2 + dr^2 + e^{B(t,r)} d\theta^2 + e^{C(t,r)} dz^2, \qquad (21)$$

where $A(r) = \ln U^2(r) - 2c_6 \int V(r)dr$,

$B(t,r) = (c_4 t + c_5)e^{-c_6\int V(r)dr} + \ln U^2(r) - 2c_2\int V(r)dr - c_3\int V(r)e^{-c_6\int V(r)dr} dr$,

$C(t,r) = (c_4 t + c_5)e^{-c_6\int V(r)dr} + \ln U^2(r) - 2c_1\int V(r)dr - c_3\int V(r)e^{-c_6\int V(r)dr} dr$,

where $V(r) = \frac{1}{U(r)}$ and $U(r) = \frac{1}{2}\int \psi(r)dr + c_5$, where $c_1, c_2, c_3, c_4, c_5, c_6, c_7 \in R(c_4 \neq 0, c_6 \neq 0, c_1 \neq c_2, c_3 = c_7 c_4 - c_6 c_5)$. The sub case when $c_1 = c_2$ will be consider later. Conformal vector fields in this case are

$$X = (c_6 t + c_7, U(r)c_{11}, \theta c_2 + c_{10}, z c_1 + c_9), \qquad (22)$$

where $c_9, c_{10}, c_{11} \in R$. Conformal factor in this case is $\psi(r) = 2\frac{d}{dr}U(r)$. In this case the above space-time (21) admits three independent conformal vector fields in which one is proper conformal and two are independent Killing vector fields. It is important to tell the readers that $c_1, c_2, c_6, c_7$ are the part of the space-times hence they can not be the generators. Generators in this case are

$$X_1 = U(r)\frac{\partial}{\partial r}, \quad X_2 = \frac{\partial}{\partial \theta}, \quad X_3 = \frac{\partial}{\partial z}.$$

Here all the structure constants are zero that is $[X_b, X_c] = 0$, for all $b, c = 1, 2, 3$. In this case Lie algebra is closed. Proper conformal vector field after subtracting Killing vector fields from equation (22) is

$$X = (c_6 t + c_8, U(r)c_{11}, \theta c_2, z c_1). \tag{23}$$

Now consider the sub case when $c_1 = c_2$ the space-time (21) becomes

$$ds^2 = -e^{A(r)} dt^2 + dr^2 + e^{B(t,r)}(d\theta^2 + dz^2). \tag{24}$$

The above space-time (24) becomes special class of non static plane symmetric space-time and admits four independent conformal vector fields in which three independent Killing vector fields are $\frac{\partial}{\partial \theta}, \frac{\partial}{\partial z}, \theta \frac{\partial}{\partial z} - z \frac{\partial}{\partial \theta}$ and one proper conformal vector field is given in equation (23). The generators in this case are

$$X_1 = U(r)\frac{\partial}{\partial r}, \quad X_2 = \frac{\partial}{\partial \theta}, \quad X_3 = \frac{\partial}{\partial z}, \quad X_4 = \theta \frac{\partial}{\partial z} - z \frac{\partial}{\partial \theta}.$$

Here the non zero components of the Lie brackets are $[X_2, X_4] = X_3$ and $[X_3, X_4] = -X_2$. In this case Lie algebra is closed.

**Case (3)**

In this case the space-time (4) becomes static cylindrically symmetric

$$ds^2 = -e^{A(r)} dt^2 + dr^2 + e^{B(r)} d\theta^2 + e^{C(r)} dz^2, \tag{25}$$

where $A(r) = \ln U^2(r) - 2c_4 \int V(r)dr,$ $\qquad B(r) = \ln U^2(r) - 2c_3 \int V(r)dr,$

$C(r) = \ln U^2(r) - 2c_2 \int V(r)dr,$ $\qquad V(r) = \frac{1}{U(r)},$ $\qquad U(r) = \frac{1}{2}\int \psi(r)dr + c_1$ and

$c_1, c_2, c_3, c_4 \in R(c_2 \neq c_3, c_2 \neq c_4, c_3 \neq c_4)$. Conformal vector fields in this case are

$$X = (c_4 t + c_8, U(r)c_{11}, \theta c_3 + c_{10}, z c_2 + c_9), \tag{26}$$

where $c_8, c_9, c_{10}, c_{11} \in R$. Conformal factor in this case is $\psi(r) = 2\frac{d}{dr}U(r)$. The above space-time (25) becomes static cylindrically symmetric space-time and admits four independent conformal vector fields in which three independent Killing vector fields are $\frac{\partial}{\partial t}, \frac{\partial}{\partial z}, \frac{\partial}{\partial \theta}$ and one proper conformal vector field. It is important to tell the readers that $c_2, c_3, c_4$ are the part of the space-times hence they can not be the generators. The generators in this case are

$$X_1 = U(r)\frac{\partial}{\partial r}, \quad X_2 = \frac{\partial}{\partial \theta}, \quad X_3 = \frac{\partial}{\partial z}, \quad X_4 = \frac{\partial}{\partial t}.$$

Here all the structure constants are zero that is $[X_b, X_c] = 0,$ for all $b, c = 1, 2, 3, 4$. In this case Lie algebra is closed. Proper conformal vector field after subtracting Killing vector fields from equation (26) is

$$X = (tc_4, U(r)c_{11}, \theta c_3, z c_2), \tag{27}$$

Now consider the sub case when $c_2 = c_3$ the space-time (25) becomes

$$ds^2 = -e^{A(r)} dt^2 + dr^2 + e^{B(r)}(d\theta^2 + dz^2). \tag{28}$$

In this case the above space-times become static plane symmetric space-time and admits four independent Killing vector fields are $\frac{\partial}{\partial t}, \frac{\partial}{\partial \theta}, \frac{\partial}{\partial z}, \theta\frac{\partial}{\partial z} - z\frac{\partial}{\partial \theta}$. In [9] the authors failed to find proper conformal vector field for the above space-times (28). In this case the above space-time (28) admits five conformal vector fields in which four are Killing vector fields which are given above and one proper conformal vector field is

$$X = (tc_4, U(r), \theta c_2, z c_2), \tag{29}$$

Conformal factor in this case is $\psi(r) = 2\frac{d}{dr}U(r)$. It is important to remind the readers that $c_2, c_4$ are the part of the space-times hence they can not be the generators. The generators in this case are

$$X_1 = U(r)\frac{\partial}{\partial r}, \ X_2 = \frac{\partial}{\partial \theta}, \ X_3 = \frac{\partial}{\partial z}, \ X_4 = \theta\frac{\partial}{\partial z} - z\frac{\partial}{\partial \theta}, \ X_5 = \frac{\partial}{\partial t}.$$

Here the non zero components of the Lie brackets are $[X_2, X_4] = X_3$ and $[X_3, X_4] = -X_2$. In this case Lie algebra is closed. The sub cases when $c_2 = c_4$ and $c_3 = c_4$ are exactly the same.